\documentclass{iaus}
\usepackage{graphicx}

\title[UV-upturn] 
{A Binary Model for the UV-upturn of Elliptical Galaxies}

\author[Han {\it et al.}]   
{Z. Han$^1$, Ph. Podsiadlowski$^2$, A.E. Lynas-Gray$^2$ \and K. Schawinski$^2$}  
\affiliation{$^1$National Astronomical Observatories / Yunnan Observatory,
\break the Chinese Academy of Sciences, Kunming 650011, China
\break email: zhanwenhan@hotmail.com\\ [\affilskip]
$^2$University of Oxford, Department of Astrophysics, Keble Road, Oxford OX1 3RH}

\pubyear{2006}
\volume{241}  
\pagerange{119--126}
\date{?? and in revised form ??}
\jname{Stellar Populations as Building Blocks of Galaxies}
\editors{A. Vazdekis \& R. Peletier, eds.}

\begin{document}

\maketitle

\begin{abstract}
The discovery of an excess of light in the far-ultraviolet (UV) in 1969 in 
elliptical galaxies was a major surprise. While it is now clear that this 
UV excess (UV-upturn) is probably caused by an old population of 
helium-burning stars. 
Han et al (2002, 2003) proposed a binary model for the formation of 
hot subdwarfs (helium burning stars) and the model can reproduce the 
observations in our Galaxy. By applying the binary model to the study 
of evolutionary population synthesis, we have obtained 
an {\it a priori} model for the 
UV-upturn of elliptical galaxies. The model shows that 
the UV-upturn is most likely resulted from binary 
interactions and it is universal (not very much metallicity-dependant) 
in ellipticals.  
This has major implications for understanding the evolution 
of the UV-upturn and elliptical galaxies in general; contrary to previous 
postulates, it implies 
that the UV-upturn is not a sign of age, but could be a potentially powerful 
indicator for a recent minor burst of star-forming activity.

\end{abstract}

A long-standing problem in the study of elliptical galaxies is 
the far-ultraviolet (UV) excess (or UV-upturn) in their spectra. 
It is now clear that UV-upturn is caused by an
old population of helium-burning stars or their descendants with a
characteristic surface temperature of 25,000\,K (\cite{Ferguson91}), also
known as hot subdwarfs. 
The origin for the population of these hot, blue stars in an otherwise
red population has, however, remained a major mystery. 
Two scenarios,
referred to as the high- and the low-metallicity scenario, have been
advanced (\cite{Yi97}; \cite{Lee94}), 
though these scenarios are quite {\it ad hoc}.
Both of these scenarios ignore the effects of binary evolution.
On the other hand, hot subdwarfs have long been studied in our own
Galaxy, and it is now well established that the vast
majority of (and quite possibly all) Galactic hot
subdwarfs are the results of binary interactions, where a star loses
all of its envelope near the tip of the red-giant branch by mass
transfer to a companion star or ejection in a common-envelope phase,
or where two helium white dwarfs merge with a combined mass
larger than $\sim 0.35\,M_{\odot}$ (see Han et al. 2002, 2003 
for references and details). In all of these cases, the remnant star 
{\it ignites helium} and becomes a hot subdwarf. 
{\it The key feature of these binary
channels is that they provide the missing physical mechanism for
ejecting the envelope and producing a hot subdwarf.  Moreover, since
it is known that these hot subdwarfs provide an important source of
far-UV light in our own Galaxy, it is reasonable to assume that they
also contribute to the far-UV in elliptical galaxies}.

To model the effects of binary evolution on the spectral appearance of
elliptical galaxies, we have performed population synthesis
study of galaxies that includes binary evolution. 
{\it It is based on a binary population synthesis model}
(\cite{Han02}; \cite{Han03}) that has been calibrated to reproduce the
short-period hot subdwarf binaries in our own Galaxy that make up the
majority of Galactic hot subdwarfs. The population
synthesis model follows the detailed time evolution of both single and
binary stars, including all binary interactions, and is capable of
simulating galaxies of arbitrary complexity, provided the
star-formation history is specified.

To obtain galaxy colours and
spectra, we have calculated spectra for hot subdwarfs using
plane-parallel static model stellar atmospheres computed with the
{\scriptsize ATLAS9} (\cite{Kurucz92}) stellar atmosphere code.
For the spectrum and colours of other single stars, we use the comprehensive
BaSeL library of theoretical stellar spectra
(\cite{Lejeune97}; \cite{Lejeune98}).

\begin{figure}
\resizebox{7cm}{!}{\includegraphics{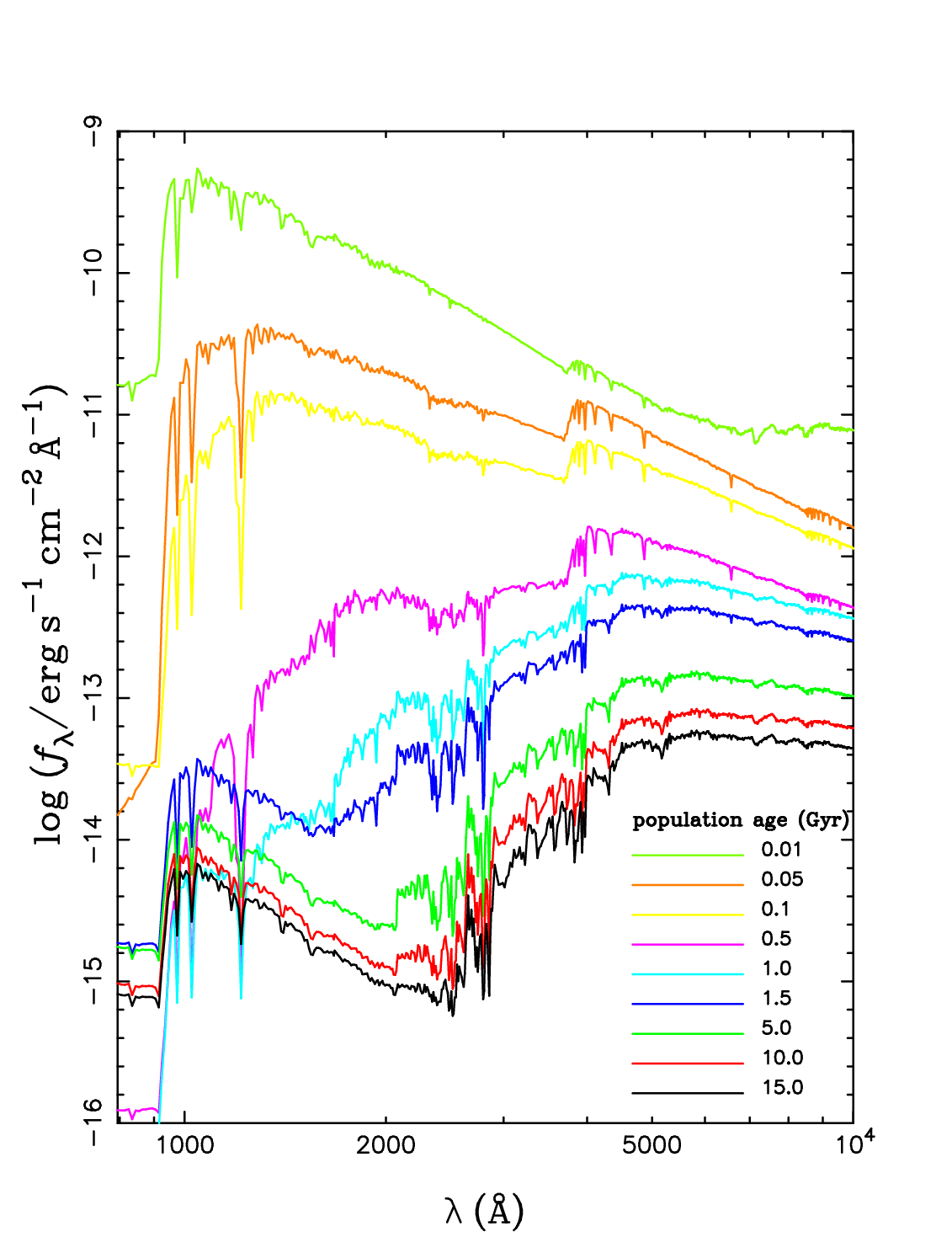}}
\resizebox{7cm}{!}{\includegraphics{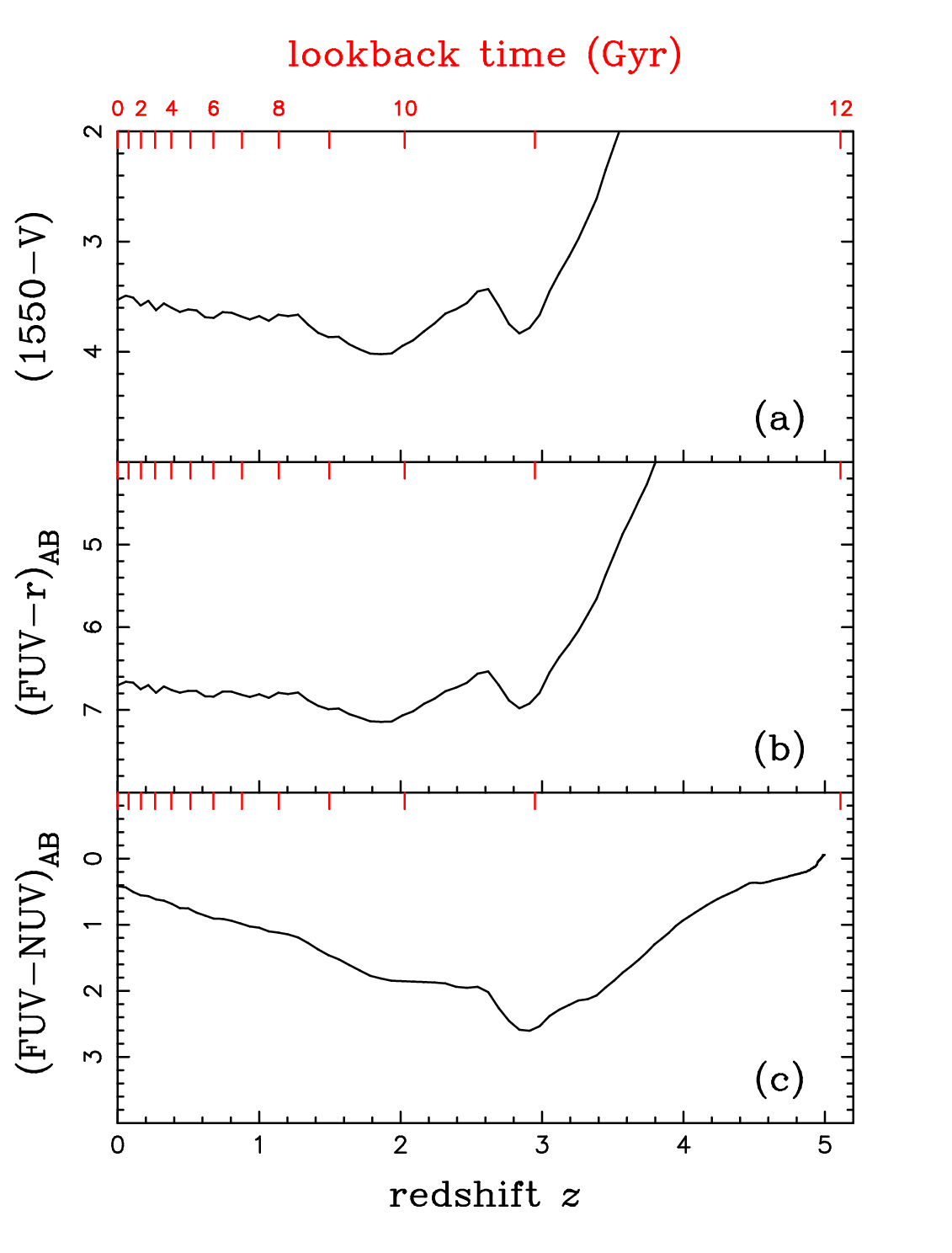}}
Figure 1: {\bf Left panel:} 
the evolution of SED (restframe and intrinsic)
for a simple stellar population (including binaries)
of $10^{10}M_\odot$ at a distance of 10Mpc.
We see that hot subdwarfs originating from binary interactions
dominate the FUV after $\sim 1$Gyr.
{\bf Right panel:}
The evolution of UV-upturn (restframe and intrinsic) with redshift 
(lookback time) for a simple stellar population (including binaries).
FUV, NUV are {\it GALEX} colours, r is a {\it SDSS} colour, 
AB means in AB magnitude.  
We adopted a star-formation redshift of $z_{\rm f}=5$, 
and cosmological parameters of
$H_0=72{\rm km/s/Mpc}$, $\Omega_{\rm M}=0.3$ and $\Omega_\Lambda=0.7$.
\end{figure}

Fig. 1 shows part of our results, and a detailed version of the paper
will be available soon.

\end{document}